\documentclass[]{pasj02} 
\usepackage[switch,mathlines]{lineno} 
\usepackage{xcolor}
\usepackage[normalem]{ulem}
\usepackage{url}

\jyear{2025}
\Received{}
\Accepted{}

\begin{document} 

\title{The line modulations of H-like Fe, Ca, Ar, and S observed with $XRISM$/Resolve in Cyg X-3 \thanks{%
This is a pre-copyedited, author-produced version of an
article accepted for publication in ``Publications of the
Astronomical Society of Japan'' following peer review. The
version of record (PASJ, 2026, Vol. xx, p. xxx) is available
online at \protect\url{[https://doi.org/10.1093/pasj/psag050]}.%
}}

\author{
Tomohiro \textsc{Hakamata}\altaffilmark{1}\altaffilmark{*}, 
Hirokazu \textsc{Odaka}\altaffilmark{1,2,3}, 
Ryota \textsc{Tomaru}\altaffilmark{1}, 
Hironori \textsc{Matsumoto}\altaffilmark{1,2}, 
Taiki \textsc{Kawamuro}\altaffilmark{1}, 
Ralf \textsc{Ballhausen}\altaffilmark{4,5,6}, 
Tim \textsc{Kallman}\altaffilmark{5}, 
Daiki \textsc{Miura}\altaffilmark{7,8}, 
Hiroya \textsc{Yamaguchi}\altaffilmark{8,7,9}, 
Teruaki \textsc{Enoto}\altaffilmark{10},
Natalie \textsc{Hell}\altaffilmark{11},
Shunji \textsc{Kitamoto}\altaffilmark{12},
Hiroshi \textsc{Nakajima}\altaffilmark{13},
Shin \textsc{Watanabe}\altaffilmark{8},
Shinya \textsc{Yamada}\altaffilmark{12},
and Kazutaka \textsc{Yamaoka}\altaffilmark{14}
}

\altaffiltext{1}{Department of Earth and Space Science, Graduate School of Science, Osaka University, 1-1 Machikaneyama, Toyonaka, Osaka 560-0043, Japan}
\altaffiltext{2}{Forefront Research Center, Graduate School of Science, Osaka University, 1-1 Machikaneyama, Toyonaka, Osaka 560-0043, Japan}
\altaffiltext{3}{Kavli Institute for the Physics and Mathematics of the Universe, The University of Tokyo, 5-1-5 Kashiwanoha, Kashiwa, Chiba, 277-8583, Japan}
\altaffiltext{4}{University of Maryland College Park, Department of Astronomy, College Park, MD 20742, USA}
\altaffiltext{5}{NASA Goddard Space Flight Center, 8800 Greenbelt Rd, Greenbelt, MD 20771, USA}
\altaffiltext{6}{Center for Research and Exploration in Space Science and Technology, NASA/GSFC (CRESST II) , Greenbelt, MD 20771, USA}
\altaffiltext{7}{Department of Physics, Graduate School of Science, The University of Tokyo, 7-3-1 Hongo, Bunkyo-ku, Tokyo 113-0033, Japan}
\altaffiltext{8}{Institute of Space and Astronautical Science (ISAS) , Japan Aerospace Exploration Agency (JAXA) , 3-1-1 Yoshinodai, Chuo-ku, Sagamihara, Kanagawa 252-5210, Japan}
\altaffiltext{9}{Department of Science and Engineering, Graduate School of Science and Engineering, Aoyama Gakuin University, 5-10-1 Fuchinobe, Chuo-ku, Sagamihara, Kanagawa 252-5258, Japan}
\altaffiltext{10}{Department of Physics, Kyoto University, Kitashirakawa-Oiwake-cho, Sakyo-ku, Kyoto, Kyoto 606-8502, Japan}
\altaffiltext{11}{Lawrence Livermore National Laboratory, 7000 East Avenue, Livermore, CA 94550, USA}
\altaffiltext{12}{Department of Physics, Rikkyo University, 3-34-1 Nishi-Ikebukuro, Toshima-ku, Tokyo 171-8501, Japan}
\altaffiltext{13}{College of Science and Engineering, Kanto Gakuin University, 1-50-1 Mutsuura Higashi, Kanazawa-ku, Yokohama, Kanagawa, 236-8501,
Japan}
\altaffiltext{14}{Institute for Space-Earth Environmental Research (ISEE) , Nagoya University, Furo-cho, Chikusa-ku, Nagoya, Aichi, 464-8601, Japan}

\email{hakamata@ess.sci.osaka-u.ac.jp}


\maketitle

\begin{abstract}
Cygnus X-3, hosting a Wolf-Rayet (WR) star whose dense wind produces various spectral lines due to photoionization by X-rays from a compact object, provides an ideal laboratory for studying wind dynamics and density structure. We measured the orbital modulations of the Fe, Ca, Ar, and S Ly$\alpha$ lines observed with the X-ray microcalorimeter (Resolve) onboard the $XRISM$, taking account of both emission and absorption lines of the Ly$\alpha$ complexes. The modulations of Doppler shifts of the Fe, Ca, Ar, and S Ly$\alpha$ lines showed amplitudes of 500 km s$^{-1}$ and phase offsets of 0.04, 0.09, 0.11, and 0.17, respectively, in units of an orbital period (4.8 hours) relative to the orbital motion of the compact object. This result indicated that H-like Fe most closely follows the compact object's motion. The line widths ranged from 400 to 1000 km s$^{-1}$. The intensities of both emission and absorption lines reached their minima around orbital phase 0.0 and their maxima around phase 0.5. The absorption peaks, however, did not align exactly with phase 0.5, suggesting the inhomogeneous structures such as an accretion wake and/or a bow shock. We compared the observed modulations with calculations based on a stellar wind model, accelerated by ultraviolet radiation from the WR star. One of the calculations qualitatively reproduced the observed trend that H-like Fe ions were concentrated near the compact object, whereas H-like S was distributed across the binary system, with H-like Ca and Ar showing intermediate spatial distributions. From this comparison, we estimated that a mass-loss rate of the stellar wind was approximately $5 \times 10^{-6}$--$1 \times 10^{-5}\ M_{\odot}\ {\rm yr}^{-1}$.

\end{abstract}


\section{Introduction}
Cygnus X-3 (Cyg X-3) is a high-mass X-ray binary (HMXB) located at a distance of 9.7 kpc (Reid \& Miller-Jones 2023), consisting of a compact object--either a black hole or a neutron star--and a WN4-6-type Wolf-Rayet (WR) star (Koljonen \& Maccarone 2017). The WR star emits a stellar wind that is denser than those from other massive stars, such as OB stars, and the wind is photoionized by X-rays from the compact object. The short orbital period of 17252 seconds ($\sim$ 4.8 hours; XRISM Collaboration et al. 2024; Kallman et al. 2019) suggests that the compact object is located in the vicinity of the WR star, deeply embedded within its dense stellar wind. It means that strong interactions between X-rays and the surrounding photoionized plasma give rise to a rich spectrum of emission and absorption lines from highly ionized elements ranging from Mg to Fe (XRISM Collaboration et al. 2024; Kallman et al. 2019).  The detection of lines from a variety of ions indicates that information on plasmas with different degree of ionization can be obtained. Since the degree of ionization depends on the distance from the compact object, each ion line can be regarded as a probe of different spatial regions within the HMXB system. 
Measuring the orbital modulation of line Doppler shifts, widths, and intensities allows us to understand the three-dimensional stellar wind structure in the HMXB system. Cyg X-3, which provides the information on a variety of lines, is the suitable target for probing the dynamics and spatial distribution of the photoionized stellar wind plasma over a wide region of the HMXB system. \par

The measurements of the Doppler shifts, widths, and intensities of the lines  require instruments with high spectral resolution. $Chandra$ is one of the few X-ray observatories capable of high-resolution spectroscopy, and it has successfully detected orbital modulation of the lines from H-like Si, S, Ar, Ca, and Fe (Vilhu \& Koljonen 2025; Vilhu et al. 2009; Stark \& Saia 2003).  However, the absorption lines of H-like ions were not detected or very weak due to the limited effective area and insufficient energy resolution of $Chandra$. Currently, measurements including absorption lines have been performed only for H-like Si and S using $Chandra$ observation data. Therefore, the motion and density distribution of ions in relatively highly ionized environments, such as H-like Ar and ions with higher atomic numbers, remain unclear. \par

The X-ray Imaging and Spectroscopy Mission ($XRISM$; Tashiro et al. 2025) addresses the limitations of effective area and energy resolution in the high-energy band. $XRISM$ was launched in 2023 September, and carries two sets of X-ray telescopes, Xtend and Resolve. Resolve is equipped with an X-ray microcalorimeter, which achieves the highest energy resolution of 4.5 eV at 6 keV among all currently launched X-ray observatories. $XRISM$ observed Cyg X-3 in March 2024, and XRISM Collaboration et al. (2024) subsequently measured the orbital modulations in the Doppler shifts of emission and absorption lines from neutral and ionized Fe in the Fe K region (6-9 keV). Their study advanced the understanding of the average motion and spatial distribution of the stellar wind within the Cyg X-3 system. Miura et al. (2025) focused on the H-like Fe located near the compact object for estimating the compact object mass, and measured the modulation of the H-like Fe emission and absorption line spectra. Their study revealed that the velocity of H-like Fe is significantly higher than the velocity of the lines of the He-like Fe, which primarily determine the shifts of the lines in the Fe K region. As indicated by these previous studies, approaches using $XRISM$ observation data have been conducted only with Fe lines. Measurements using lines from a variety of elements other than Fe are needed. \par

It is effective to measure spectral lines from ions with different photoionization parameters to investigate the dynamics and spatial distribution of photoionized plasma across a broad range of HMXB systems. In XRISM Collaboration et al. (2024), they measured the lines from different ionization states of the same element.  However, the Fe K region contains numerous lines, making its spectral structure extremely complex. For example, the lines from lower-ionization Fe ions such as Li- and Be-like Fe can contaminate the lines of He-like Fe. This method may lead to the uncertainties in determining plasma dynamics and spatial distribution becoming large. Instead, we measured the lines from H-like ions of different elements. The spectral lines from H-like ions are largely free from contamination by other lines. In addition, the intensity ratio of the Ly$\alpha_1$ and Ly$\alpha_2$ lines from H-like ions is determined only by atomic physics and is approximately 2:1. Therefore, measuring H-like ion lines allows us to extract information on the dynamics and spatial distribution of the plasma with smaller uncertainties. \par

In this work, we measured the orbital modulations in the Doppler shifts, widths, and intensities of the lines from H-like ions in the $XRISM$ spectra using models that include absorption lines. We aim to elucidate the motion and spatial distribution of the stellar wind influenced by the WR star and the compact object throughout the entire Cyg X-3 system. To this end, we performed a comprehensive analysis of the modulations of H-like Fe, Ca, Ar, and S. Additionally, we calculated the distribution of ion densities resulting from photoionization based on a simple stellar wind model as in Watanabe et al. (2006), who computed the ionization structure of the Vela X-1 wind. This calculation requires the spectrum and luminosity of the compact object serving as the ionizing source. The spectrum of the ionizing source is required up to the hard X-ray band above $\sim$ 10 keV. Accordingly, we also analyzed a simultaneous observation from the Nuclear Spectroscopic Telescope Array ($NuSTAR$; Harrison et al. 2013). \par

This paper is organized as follows: In Section \ref{sec:obsredct}, we describe the observational data from $XRISM$, as well as their data reduction. In Section \ref{sec:ana}, we present the analysis results of the H-like ion line spectra obtained with $XRISM$. In Section \ref{sec:disc}, we discuss the stellar wind structure by comparing our measurements with the previous studies and the results of photoionization equilibrium plasma calculations. Finally, in Section \ref{sec:conc}, we give the conclusions of our study. In Appendix, we describe the $NuSTAR$ observations and data reduction, and present the results of the broadband spectral analysis.\par


\section{Observation and data reduction}
\label{sec:obsredct}

Cyg X-3 was observed by $XRISM$ during the Performance Verification (PV) phase from 2024 March 24 06:53:50 TT to 2024 March 25 15:37:46 TT (ObsID: 300065010). 
During the observation, the compact object was in the "hypersoft state" before the radio burst (XRISM Collaboration et al. 2024). The observation resulted in a net exposure of 66.6 ks and covering seven orbital periods. \par

\begin{figure*}[!]
 \begin{minipage}{0.5\hsize}
\includegraphics[keepaspectratio,scale=1.3,angle=0]{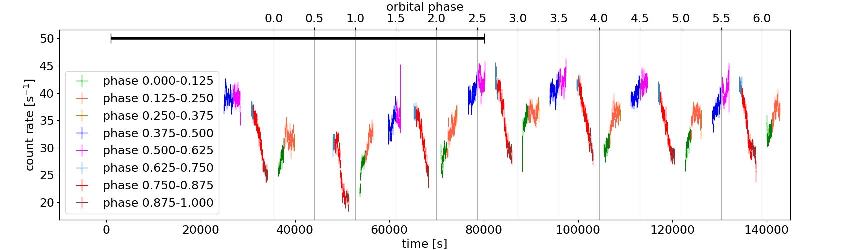}
 \end{minipage}
 \caption{Light curve in the 2--12 keV band observed by $XRISM$/Resolve. The selected event grade is Hp. The lower horizontal axis shows the elapsed time in seconds from MJD 60393.
The upper horizontal axis shows the orbital phase, which increases by +1 for each period. The green, orange, yellow, blue, magenta, skyblue, red, and brown crosses represent the observed count rates in phase 0.000--0.125, 0.125--0.250, 0.250--0.375, 0.375--0.500, 0.500--0.625, 0.625--0.750, 0.750--0.875, and 0.875--1.000, respectively. The data gaps around orbital phases 0.0, 0.4, and 0.5 are due to the SAA and earth occultation. The black bar represents the timing of the $NuSTAR$ observation. Note that the true difference between the maximum and minimum count rates may be larger than that indicated by this light curve since the grade branching ratio into Hp varies, with a time-averaged value of 43 \% and a standard deviation of 10 \%.}\label{fig:xrism_lc}
\end{figure*}

We divided the orbital phase into eight bins and extracted a spectrum for each phase. 
We used the orbital period of 17,252 seconds and the MJD of 60,393.8100 corresponding to the orbital phase of 0.0. We defined the superior conjunction (the compact object is behind the WR star) as orbital phase 0.0, and the inferior conjunction (the compact object is in front of the WR star) as orbital phase 0.5.
The analysis phase bins were defined as follows: 0.000--0.125, 0.125--0.250, 0.250--0.375, 0.375--0.500, 0.500--0.625, 0.625--0.750, 0.750--0.875, and 0.875--1.000.
 Figure \ref{fig:xrism_lc} shows the light curve observed by $XRISM$/Resolve with each defined phase interval. 
 Due to the overlap between the periods of South Atlantic Anomaly (SAA) and earth occultation with the orbital period of Cyg X-3, there are periods during which observations could not be conducted around orbital phases 0.0, 0.4, and 0.5.   \par
 
In Figure \ref{fig:xrism_lc}, the correction for the grade branching ratio of Hp events was not applied.
The true flux variation is larger than that shown in Figure \ref{fig:xrism_lc} since the branching ratio into the Hp grade varies, with a time-averaged value of 43 \% and a standard deviation of 10 \%.
The average flux at each orbital phase varies over the orbital cycle.
This variation is estimated to be less than 20\% and does not significantly affect the analysis in Section \ref{sec:ana}, where its main potential impact would be on the measurement of the line flux modulation.
 \par

We used data and tools provided by XRISM Science Data Center (SDC) at NASA Goddard Space Flight Center for the analysis.
We used the cleaned event file reprocessed by the SDC on 2024 July 24. 
In the cleaned event file, events with STATUS bits 2, 3, and 6 set were excluded using \texttt{rslflagpix} (the detailed screening logs applied in producing the cleaned event file are provided in \texttt{xa300065010\_joblog.html}). No additional event screening was applied beyond this standard processing.
We extracted spectra of only High primary (Hp) events excluding events from pixel 27 due to unpredictable gain excursions in this pixel. We binned the spectrum at 2 eV per bin to reduce the computation time. We performed the data reduction and analysis of the Resolve data using $XRISM$ SDC tools, Build7, which are part of the non-public HEASoft framework and include XSPEC 12.13.1. The Large RMF and ARF files were generated with \texttt{rslmkrmf} and \texttt{xaarfgen}, respectively. For making the RMF and ARF files, we used CALDB 7, which is the non-public CALDB released on 2023 August 15.
Anomalous Low secondary (Ls) events may cause uncertainties in the absolute emission-line flux, possibly making discussions based on absolute flux values less robust. We focus on the relative intensity variations, rather than on the absolute flux values.\par

\section{Analysis and results}
\label{sec:ana}
\subsection{Spectral analysis}
\begin{figure*}[!]
\begin{minipage}{0.5\hsize}
\includegraphics[keepaspectratio,scale=0.53,angle=0]{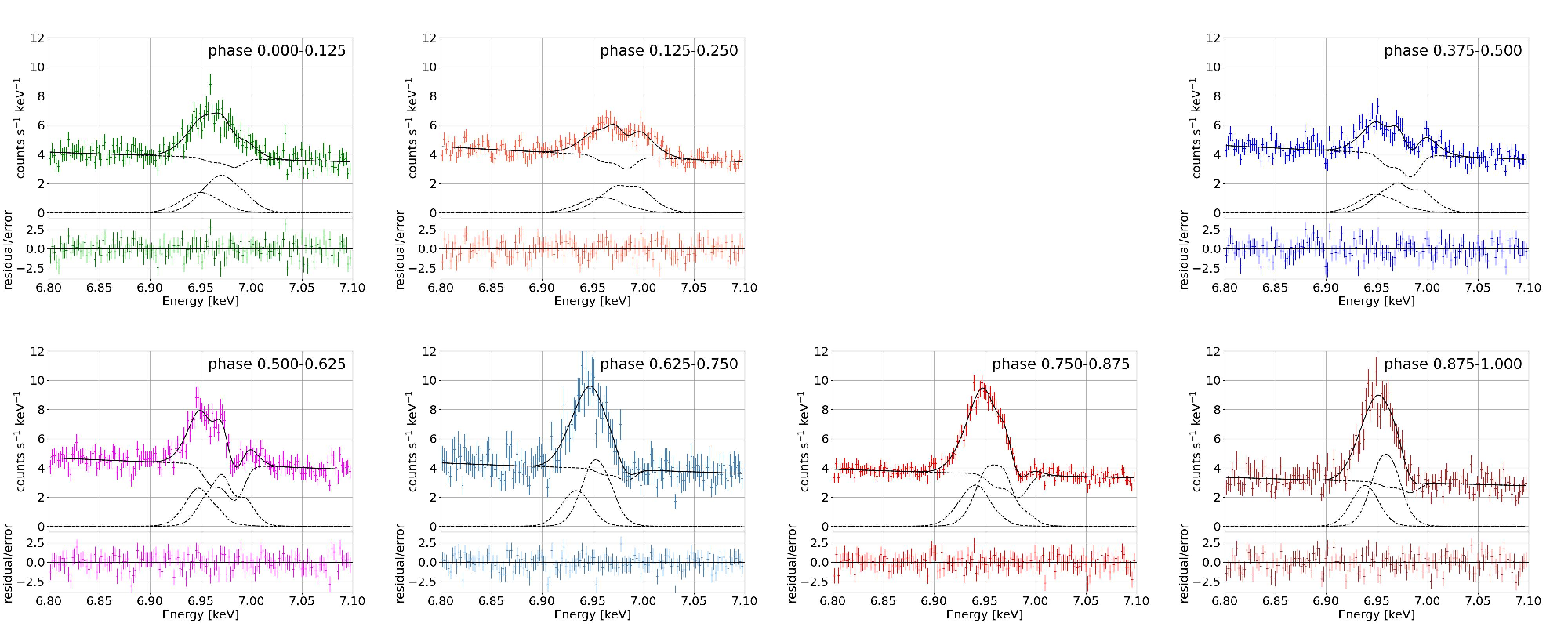}
 \end{minipage}
 \caption{Spectra of Fe Ly$\alpha$ lines. In each panel, the upper and lower parts show the observed spectrum and the residual, respectively. The crosses represent the observed spectra. The correspondence between each color and phase is the same as in Figure \ref{fig:xrism_lc}. The black solid lines represent the best-fit models. The black dashed lines represent the components of absorbed $powerlaw$, and Ly${\alpha_{1}}$ and Ly${\alpha_{2}}$ emission lines. The spectra are binned at 2 eV per bin.}
 \label{fig:xrism_spec_fe}
\end{figure*}

\begin{figure*}[!]
\begin{minipage}{0.5\hsize}
\includegraphics[keepaspectratio,scale=0.53,angle=0]{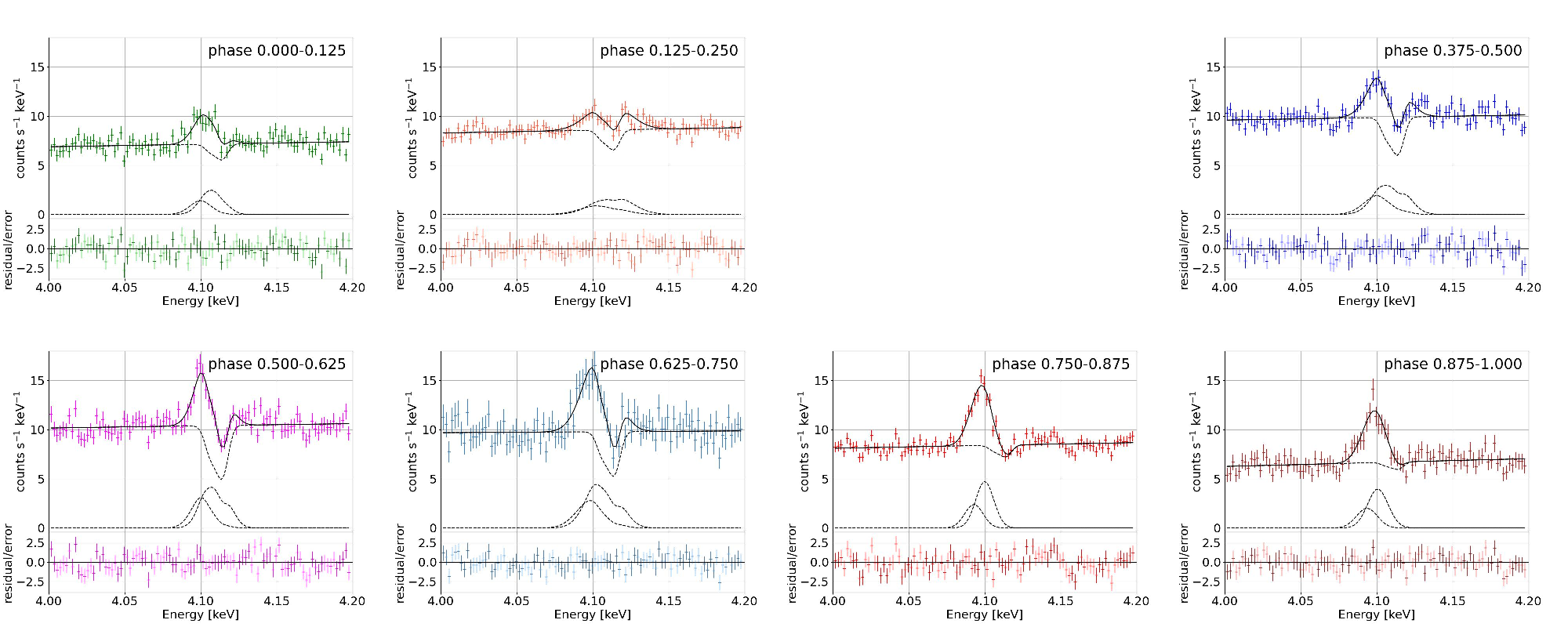}
 \end{minipage}
 \caption{Spectra of Ca Ly$\alpha$ lines, similar to those in Figure \ref{fig:xrism_spec_fe}.}
 \label{fig:xrism_spec_ca}
\end{figure*}

\begin{figure*}[!]
\begin{minipage}{0.5\hsize}
\includegraphics[keepaspectratio,scale=0.53,angle=0]{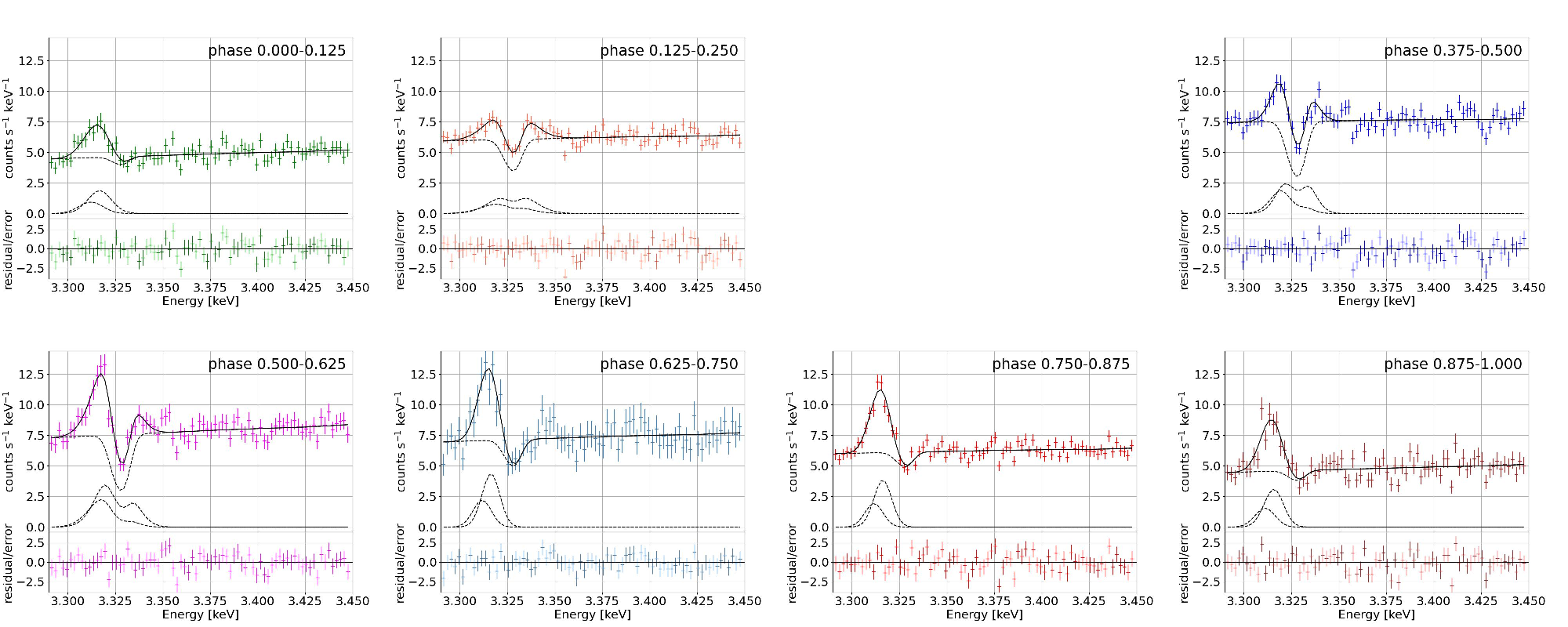}
\end{minipage}
 \caption{Spectra of Ar Ly$\alpha$ lines, similar to those in Figure \ref{fig:xrism_spec_fe}.}
 \label{fig:xrism_spec_ar}
\end{figure*}

\begin{figure*}[!]
\begin{minipage}{0.5\hsize}
\includegraphics[keepaspectratio,scale=0.53,angle=0]{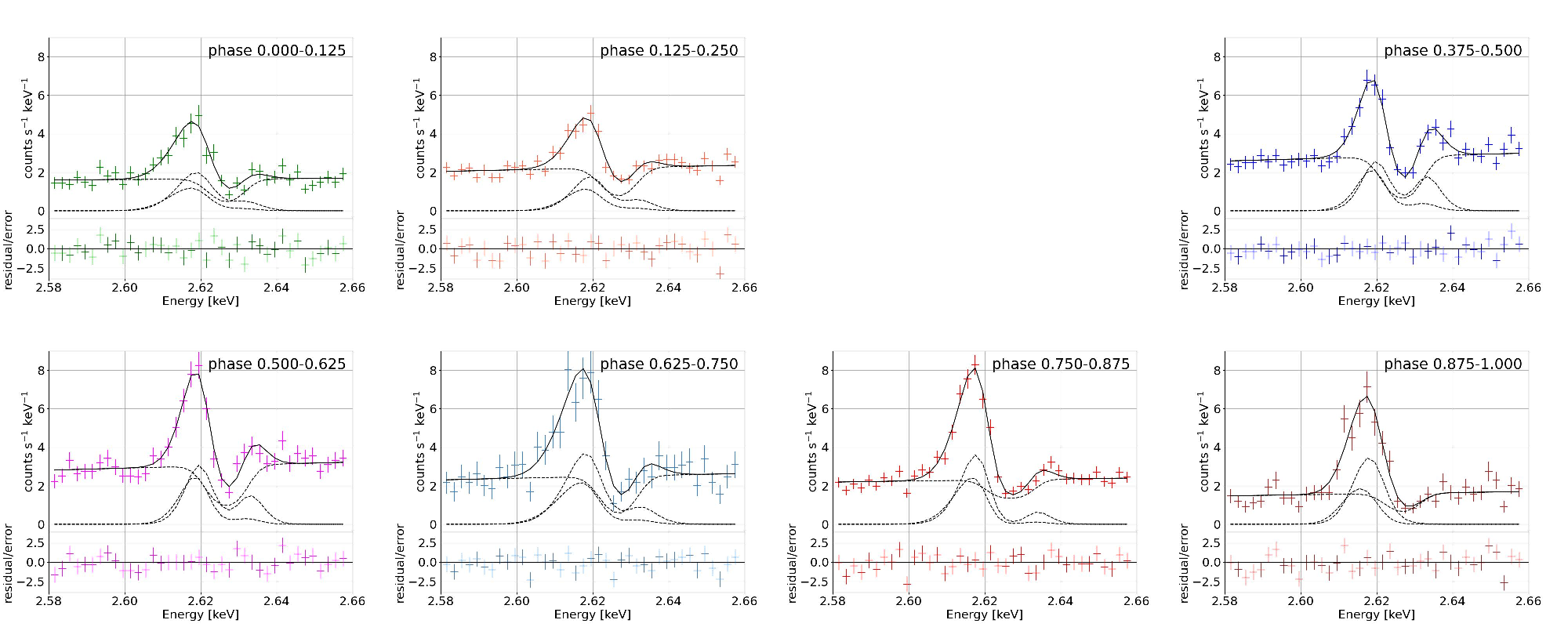}
\end{minipage}
 \caption{Spectra of S Ly$\alpha$ lines, similar to those in Figure \ref{fig:xrism_spec_fe}.}
 \label{fig:xrism_spec_s}
\end{figure*}

We measured the orbital modulations of the Ly${\alpha_{1}}$ and Ly${\alpha_{2}}$ lines from H-like Fe, Ca, Ar, and S ions. Figures \ref{fig:xrism_spec_fe}--\ref{fig:xrism_spec_s} show the spectra of the Fe, Ca, Ar, and S Ly${\alpha}$ lines extracted at each phase. The spectra of phase 0.250--0.375 were excluded because the net exposure was extremely short, leading to the insufficient photon counts for analysis. Overall, the line intensities in phase 0.5--1.0 tend to be stronger than those in phase 0.0--0.5. The Fe Ly${\alpha}$ absorption lines are often located near the center of the emission lines and do not exhibit a typical P Cygni profile. In contrast, the Ca, Ar, and S Ly${\alpha}$ lines show P Cygni-like structures, particularly in phase 0.50--0.75. On the other hand, the absorption lines appear near the centers of the emission lines in phase 0.125--0.500, similar to Fe Ly${\alpha}$. Such detailed line structures have never been resolved with $Chandra$; they were detected for the first time owing to the high spectral resolution and large effective area of $XRISM$. \par

We analyzed the Ly${\alpha_{1}}$ and Ly${\alpha_{2}}$ lines of Fe, Ca, Ar, and S ions. The spectral analysis was conducted using the 6.80--7.10 keV band for Fe, 4.00--4.20 keV for Ca, 3.29--3.45 keV for Ar, and 2.58--2.66 keV for S. 
We adopted a model that accounts for both emission and absorption features, because both were present in the spectra. 
The obtained spectra were expected to include both the Ly${\alpha_{1}}$ and Ly${\alpha_{2}}$ lines. 
We incorporated these components into the model. 
That is, the model includes two emission components and two absorption components. 
A phenomenological continuum component was also included to represent the radiation from the compact object.
The radiation from the compact object is line-absorbed by the ions in the surrounding stellar wind.
The line emission is subject to line absorption since the stellar wind contributes to both line absorption and line emission.
The continuum emission and the line emission arise from different regions. 
In principle, the absorption column is different for each component. 
Nevertheless, the photon statistics of the observed spectra do not allow us to constrain separate absorption components for the continuum and line emission. 
Accordingly, we adopt a simplified approximation in which the same line-absorption component is applied to all emission components.
Such a fitting model was also supported by XRISM Collaboration et al. (2024).
The model we used in XSPEC is given by \texttt{(powerlaw+zgauss+constant$\times$zgauss)$\times$gabs$\times$gabs}. \texttt{zgauss}, \texttt{gabs}, \texttt{powerlaw}, and \texttt{constant} represent the emission lines, absorption lines, and continuum, and intensity ratio between the Ly${\alpha_{1}}$ and Ly${\alpha_{2}}$ emission lines, respectively. \par

We simultaneously fitted the Ly$\alpha$ line spectra across all orbital phases simultaneously using the model described above.
The widths of the Ly${\alpha_{1}}$ and Ly${\alpha_{2}}$ lines (1 $\sigma$) were linked for each phase. 
The energy ratios between the Ly${\alpha_{1}}$ and Ly${\alpha_{2}}$ absorption lines were fixed to their intrinsic energy ratios to link the Doppler shifts of these lines with the same value. The Doppler shifts of the Ly${\alpha_{1}}$ and Ly${\alpha_{2}}$ emission lines were tied together. The normalization ratios of the emission lines, as well as the absorption line strengths, were fixed according to the oscillator strengths recorded in the MONACO database (Odaka et al. 2011), which was constructed by using the Flexible Atomic Code (Gu 2008). In Table \ref{tab:mncdb}, the oscillator strengths used in this paper are summarized. For the fitting process, we employed the C-statistic (Cash 1979), where a best-fit model is estimated by minimizing the log-likelihood $C$ calculated under the assumption of a Poisson distribution. \par

We fitted the central energies and intensities of the emission and absorption Ly$\alpha$ lines of Fe, Ca, Ar, and S independently for each orbital phase. The widths of the emission lines were allowed to vary independently for each phase, whereas the widths of the absorption lines were linked across all phases for stable fitting since the determination of the widths of the absorption lines in each phase would be severely limited by the low photon counts. 
The spectra in some orbital phase bins such as phase 0.0-0.125 did not show strong absorption features, unconstraining the absorption line's central energy and width, but the others with relatively strong features did not show significant central energy and width variations. 
Therefore, we assumed that the absorption line energy and width did not change throughout the binary orbit and tie those values across the spectra.
The depths of the absorption lines were allowed to vary independently, as modulations were observed in the spectra. \par

\begin{table}[!]
  \tbl{The oscillator strengths used in this paper}{%
   \begin{tabular}{ccccc}
      \hline
      \hline \\
      line ID & energy [keV] & oscillator strength \\ 
      \hline 
      Fe Ly${\alpha_{1}}$ & 6.973 & 0.273 \\
      Fe Ly${\alpha_{2}}$ & 6.952 & 0.136 \\
      Ca Ly${\alpha_{1}}$ & 4.108 & 0.275  \\
      Ca Ly${\alpha_{2}}$ & 4.100 & 0.137  \\
      Ar Ly${\alpha_{1}}$ & 3.323 & 0.275  \\
      Ar Ly${\alpha_{2}}$ & 3.318 & 0.138  \\
      S Ly${\alpha_{1}}$ & 2.623 & 0.276  \\
      S Ly${\alpha_{2}}$ & 2.620 & 0.138  \\
      \hline
    \end{tabular}}\label{tab:mncdb}
\begin{tabnote}
Note: Similar to the oscillator strength, the line energy is the value referenced from the MONACO database.\\
\end{tabnote}
\end{table}

\begin{table*}[!]
  \tbl{The spectral fitting results of Fe Ly$\alpha$ lines.}{%
  \begin{tabular}{cccccccc}
      \hline
      \hline
      orbital phase & 0.000-0.125 & 0.125-0.250 & 0.375-0.500 & 0.500-0.625 & 0.625-0.750 & 0.750-0.875 & 0.875-1.000  \\ 
      \hline
      & & & & Ly$\alpha_{1}$ emission line  & & & \\
      \hline
      velocity [km s$^{-1}$] & 34$\pm79$ & $-372\pm74$ & $-120_{-75}^{+74}$ & $-51_{-51}^{+47}$ & 784$\pm81$ & 427$\pm35$ & 558$_{-53}^{+52}$  \\
      width [km s$^{-1}$] & 828$_{-73}^{+79}$ & 914$_{-69}^{+74}$ & 884$_{-70}^{+76}$ & 684$_{-49}^{+52}$ & 626$_{-77}^{+85}$ & 730$_{-35}^{+38}$ & 589$_{-53}^{+56}$  \\
      flux [10$^{-3}$ photons cm$^{-2}$ s$^{-1}$] & 2.01$\pm0.13$ & 1.83$\pm0.11$ & 2.10$\pm0.13$ & 3.50$\pm0.18$ & 2.58$\pm0.21$ & 3.53$\pm0.12$ & 2.80$\pm0.16$  \\
      \hline
      & & & & Ly$\alpha_{1}$ absorption line  & & & \\
      \hline
      velocity [km s$^{-1}$] & & & & $-442_{-40}^{+41}$ & & &  \\
      width [km s$^{-1}$] & & & & 340$_{-21}^{+24}$ & & &  \\
      optical depth & 0.18$\pm0.05$ & 0.26$\pm0.05$ & 0.50$\pm0.06$ & 0.88$\pm0.07$ & 0.22$\pm0.10$ & 0.60$\pm0.05$ & 0.26$\pm0.08$  \\
      \hline
      $C$ (dof) & 958.73 (999) & & & & & & \\
      \hline
      \hline
    \end{tabular}}\label{tab:lyafe}
\end{table*}

\begin{table*}[!]
  \tbl{The spectral fitting results of Ca Ly$\alpha$ lines.}{%
   \begin{tabular}{cccccccc}
      \hline
      \hline
      orbital phase & 0.000-0.125 & 0.125-0.250 & 0.375-0.500 & 0.500-0.625 & 0.625-0.750 & 0.750-0.875 & 0.875-1.000  \\ 
      \hline
      & & & & Ly$\alpha_{1}$ emission line  & & & \\
      \hline
      velocity [km s$^{-1}$] & $-21\pm103$ & $-382\pm143$ & $-131_{-80}^{+81}$ & $-148_{-57}^{+59}$ & 37$_{-113}^{+115}$ & 528$\pm49$ & 476$_{-90}^{+89}$  \\
      width [km s$^{-1}$] & 518$_{-100}^{+111}$ & 915$_{-127}^{+150}$ & 693$_{-77}^{+83}$ & 532$_{-59}^{+62}$ & 708$_{-97}^{+111}$ & 425$_{-49}^{+53}$ & 490$_{-81}^{+92}$  \\
      flux [10$^{-3}$ photons cm$^{-2}$ s$^{-1}$] & 1.02$_{-0.13}^{+0.14}$ & 1.17$\pm0.14$ & 1.91$_{-0.16}^{+0.17}$ & 2.50$\pm0.20$ & 2.73$_{-0.32}^{+0.33}$ & 1.41$\pm0.10$ & 1.35$\pm0.16$  \\
      \hline
      & & & & Ly$\alpha_{1}$ absorption line  & & & \\
      \hline
      velocity [km s$^{-1}$] & & & &  $-482_{-76}^{+36}$ & & &  \\
      width [km s$^{-1}$] & & & & 227$_{-30}^{+34}$ & & &  \\
      optical depth & 0.29$_{-0.07}^{+0.08}$ & 0.30$\pm0.07$ & 0.55$\pm0.09$ & 0.83$\pm0.13$ & 0.69$\pm0.14$ & 0.18$\pm0.06$ & 0.14$\pm0.10$  \\
      \hline
      $C$ (dof) & 656.40 (649) & & & & & & \\
      \hline
      \hline
    \end{tabular}}\label{tab:lyaca}
\end{table*}

\begin{table*}[!]
  \tbl{The spectral fitting results of Ar Ly$\alpha$ lines.}{%
   \begin{tabular}{cccccccc}
      \hline
      \hline
      orbital phase & 0.000-0.125 & 0.125-0.250 & 0.375-0.500 & 0.500-0.625 & 0.625-0.750 & 0.750-0.875 & 0.875-1.000  \\ 
      \hline
      & & & & Ly$\alpha_{1}$ emission line  & & & \\
      \hline
      velocity [km s$^{-1}$] & 537$_{-137}^{+136}$ & $-436_{-145}^{+147}$ & $-402\pm68$ & $-178_{-80}^{+81}$ & 571$_{-113}^{+112}$ & 585$\pm55$ & 627$_{-104}^{+103}$  \\
      width [km s$^{-1}$] & 567$_{-104}^{+127}$ & 925$_{-128}^{+148}$ & 617$_{-69}^{+77}$ & 755$_{-75}^{+84}$ & 413$_{-99}^{+120}$ & 407$_{-52}^{+56}$ & 455$_{-95}^{+111}$ \\
      flux [10$^{-3}$ photons cm$^{-2}$ s$^{-1}$] & 0.92$\pm0.15$ & 1.40$\pm0.18$ & 2.56$_{-0.23}^{+0.24}$ & 3.21$\pm0.28$ & 1.63$_{-0.27}^{+0.29}$ & 1.41$\pm0.12$ & 1.25$_{-0.18}^{+0.19}$  \\
      \hline
      & & & & Ly$\alpha_{1}$ absorption line  & & & \\
      \hline
      velocity [km s$^{-1}$] & & & &  $-575$$_{-23}^{+61}$ & & &  \\
      width [km s$^{-1}$] & & & & 270$\pm31$ & & &  \\
      optical depth & 0.18$_{-0.11}^{+0.12}$ & 0.53$\pm0.10$ & 0.88$\pm0.14$ & 0.99$\pm0.16$ & 0.35$_{-0.14}^{+0.15}$ & 0.21$\pm0.07$ & 0.32$_{-0.20}^{+0.22}$  \\
      \hline
      $C$ (dof) & 545.26 (509) & & & & & & \\
      \hline
      \hline
    \end{tabular}}\label{tab:lyaar}
\end{table*}

\begin{table*}[!]
  \tbl{The spectral fitting results of S Ly$\alpha$ lines.}{%
   \begin{tabular}{cccccccc}
      \hline
      \hline
      orbital phase & 0.000-0.125 & 0.125-0.250 & 0.375-0.500 & 0.500-0.625 & 0.625-0.750 & 0.750-0.875 & 0.875-1.000  \\ 
      \hline
      & & & & Ly$\alpha_{1}$ emission line  & & & \\
      \hline
      velocity [km s$^{-1}$] & 81$_{-104}^{+107}$ & $-64_{-83}^{+85}$ & -360$_{-46}^{+48}$ & $-280_{-51}^{+53}$ & 112$_{-123}^{+126}$ & 19$_{-51}^{+52}$ & 444$_{-84}^{+81}$  \\
      width [km s$^{-1}$] & 743$_{-90}^{+102}$ & 703$_{-76}^{+84}$ & 622$_{-46}^{+48}$ & 610$_{-49}^{+53}$ & 807$_{-105}^{+120}$ & 706$\pm50$ & 510$_{-66}^{+79}$  \\
      flux [10$^{-3}$ photons cm$^{-2}$ s$^{-1}$] & 3.27$_{-0.36}^{+0.37}$ & 3.21$\pm0.32$ & 7.71$_{-0.62}^{+0.63}$ & 8.18$_{-0.72}^{+0.75}$ & 6.48$_{-0.80}^{+0.84}$ & 7.28$_{-0.48}^{+0.50}$ & 3.39$_{-0.36}^{+0.40}$  \\
      \hline
      & & & & Ly$\alpha_{1}$ absorption line  & & & \\
      \hline
      velocity [km s$^{-1}$] & & & &  $-576$$_{-57}^{+28}$ & & &  \\
      width [km s$^{-1}$] & & & & 353$\pm20$ & & &  \\
      optical depth & 1.07$_{-0.15}^{+0.16}$ & 1.16$\pm0.20$ & 1.86$_{-0.27}^{+0.28}$ & 1.95$_{-0.30}^{+0.31}$ & 1.35$_{-0.28}^{+0.30}$ & 3.06$_{-0.53}^{+0.57}$ & 0.85$_{-0.25}^{+0.26}$  \\
      \hline
      $C$ (dof) & 247.35 (229) & & & & & & \\
      \hline
      \hline
    \end{tabular}}\label{tab:lyas}
\end{table*}

\begin{figure*}[!]
\includegraphics[keepaspectratio,scale=0.354,angle=0]{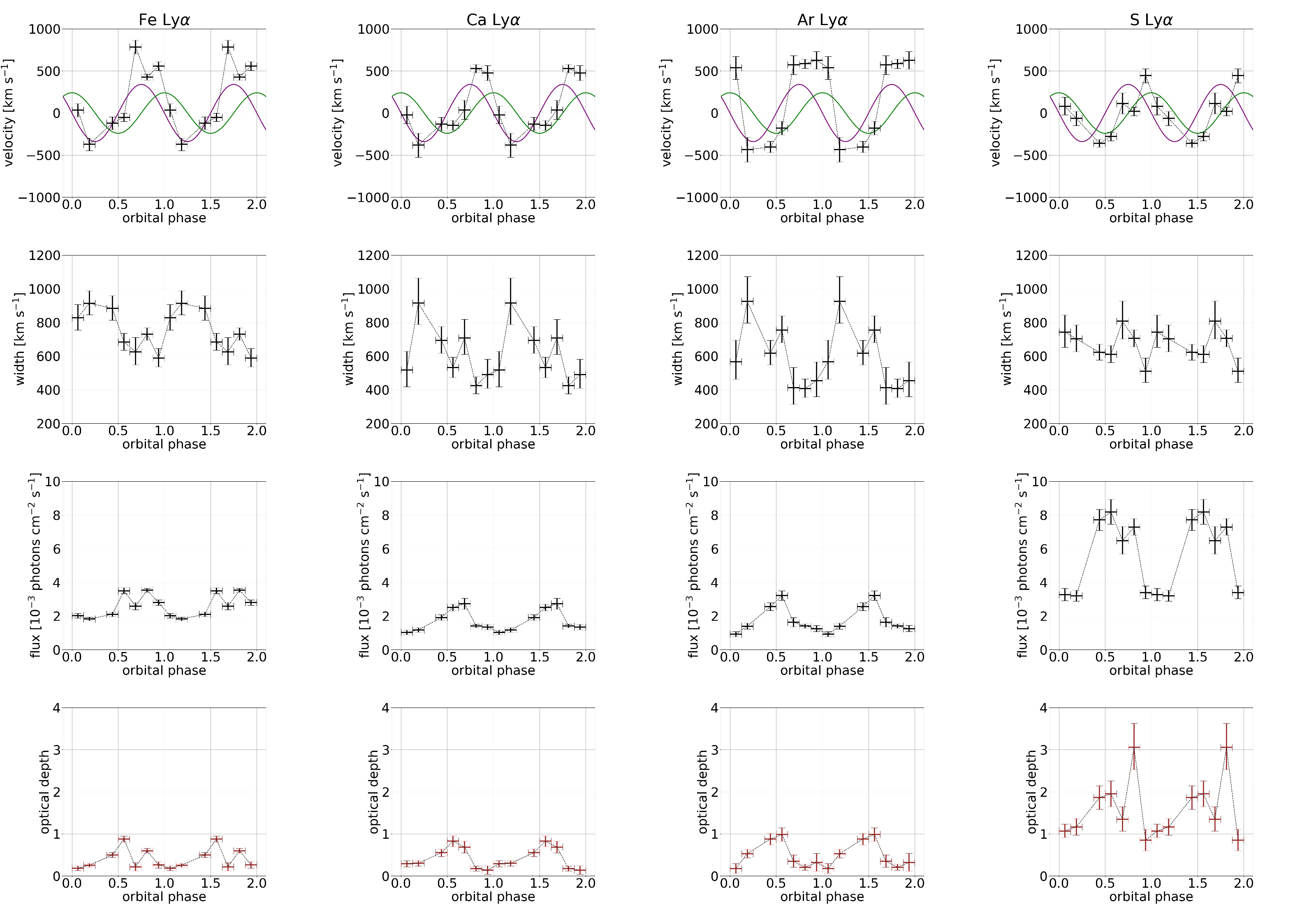}
 \caption{Orbital modulations of Ly${\alpha_{1}}$ lines. The orbital modulations of Fe, Ca, Ar, and S are shown from left to right, with those of the emission line velocities, widths, and normalizations (black), and the absorption line strengths (brown) represented from top to bottom. In each panel, the crosses and dashed lines represent the observed values. The purple and green lines in the top four panels represent the calculated orbital modulation of the compact object motion and the velocity of the stellar wind at the L$_{1}$ point on line of sight. The velocity of the stellar wind is calculated with the CAK model.}
 \label{fig:orb_mod}
\end{figure*}


We obtained reasonable fits; the best-fit models are superimposed in Figures \ref{fig:xrism_spec_fe}--\ref{fig:xrism_spec_s} and summarized in Tables \ref{tab:lyafe}--\ref{tab:lyas}. There were no significant residuals between the observed spectra and the best-fit models. In these figures and tables, the positive velocity values correspond to redshifts, while negative values correspond to blueshifts. Unless otherwise noted, all errors hereafter are quoted at a 90\% confidence level.
The orbital modulations of each parameter of the Ly${\alpha_{1}}$ line are shown in Figure \ref{fig:orb_mod}.
The modulations of velocities of the Fe Ly$\alpha$ emission lines exhibit phases more closely aligned with the calculated orbital motion of the compact object than other Ly$\alpha$ lines (for quantitative measurements, see section \ref{subsubsec:ccf}). 
The line widths for all elements exhibit the modulations over the range of 400--1000 km s$^{-1}$. 
The modulations of emission line flux and optical depth exhibit similar phases. They reach the minimum around orbital phase 0.0 and the maximum around phase 0.5. 
 \par

We checked the consistency between our analysis and results of Miura et al. (2025), who analyzed the Fe Ly$\alpha$ lines in the same observational data using the different method (Markov chain Monte Carlo method). 
No notable difference was found in the modulation of the Fe Ly$\alpha$ emission lines compared with Miura et al. (2025). 
The central energies and the widths of absorption lines in all orbital phases were consistent with Miura et al. (2025) within the 90\% confidence interval.
Although a relatively large discrepancy was only found in the widths of the Fe Ly$\alpha$ absorption lines in the orbital phase range 0.625--0.750, this discrepancy is not serious since the absorption lines are very weak in this phase. \par

\subsection{Phase difference from the orbital motion of compact object}
\label{subsubsec:ccf}
\begin{figure*}[!]
\includegraphics[keepaspectratio,scale=0.479,angle=0]{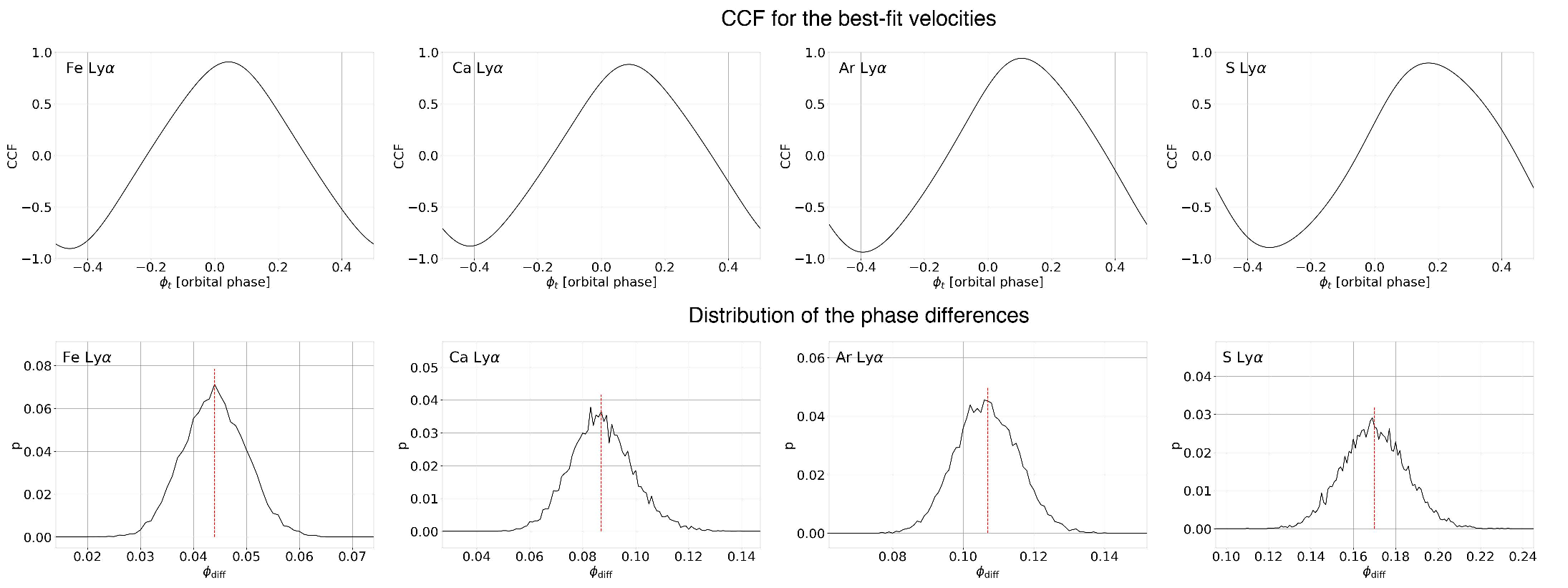}
 \caption{Analysis results of the phase difference using the CCF. In the upper four panels, the $\phi_{t}$ for Fe, Ca, Ar, and S, along with their corresponding CCF values measured from the best-fit values of the emission line velocities, are presented. In the bottom four panels, the error distributions of $\phi_{{\rm diff}}$ obtained by the Monte Carlo method were shown by the black solid lines. The red dashed lines represent the $\phi_{{\rm diff}}$ for the best-fit emission line velocities.}
 \label{fig:ccfres}
\end{figure*}

\begin{figure}[!]
\includegraphics[keepaspectratio,scale=0.833,angle=0]{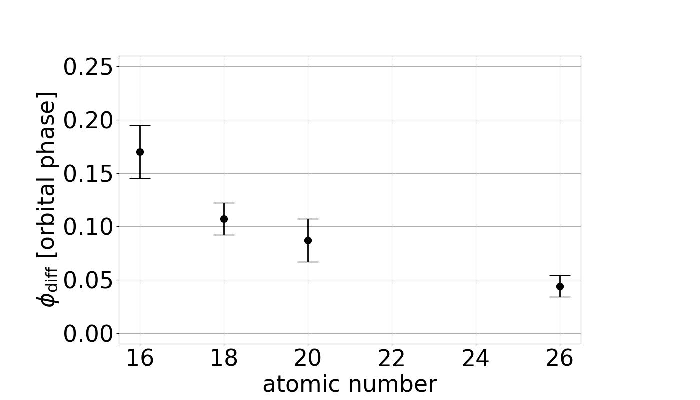}
 \caption{$\phi_{{\rm diff}}$ of each element. The confidence interval of the error bar is 90\%.}
 \label{fig:ccfpd}
\end{figure}

We measured the phase difference between the Ly${\alpha}$ emission line velocities and the motion of the compact object along the line of sight. For the analysis we used a cross-correlation function (CCF), given by 
\begin{equation}
    {\rm CCF}(\phi_{t}) = \frac{\sum_{i} (f_{1}(\phi_{i}-\phi_{t})-\bar{f_{1}})(f_{2}(\phi_{i})-\bar{f_{2}})}{\sqrt{\sum_{i} (f_{1}(\phi_{i}-\phi_{t})-\bar{f_{1}})^{2}} \sqrt{\sum_{i} (f_{2}(\phi_{i})-\bar{f_{2}})^{2}}}
\end{equation}
where $f_{1}$, $f_{2}$, $\phi_{i}$, $\phi_{t}$ represent the model of orbital modulation in the compact object's motion, the observed emission line velocity modulations, the orbital phase for each bin, and the shift of model in the calculation, respectively. The overbars denote the mean values. The model of compact object's motion was based on the following parameters: the mass of the WR star (10.3 M$_{\odot}$), the mass of the compact object (2.4 M$_{\odot}$), and the inclination angle (\timeform{29D.5}$\pm$\timeform{1D.2}). The masses and inclination angle are given by Zdziarski et al. (2013) and Antokhin et al. (2022), respectively. We assumed the circular orbit because the orbital eccentricity is small, at 0.03 (Antokhin \& Cherepashchuk 2019).  
The calculated line-of-sight velocity modulation of the orbital motion of the compact object is shown with a purple line in Figure \ref{fig:orb_mod}.
We defined $\phi_{{\rm diff}}$ as the value of $\phi_{t}$ at which the CCF reaches its maximum, referred to as the phase difference. We estimated the error in the phase difference using the Monte Carlo method. Based on the error distribution of each emission line velocity, we generated 10,000 modulations and measured $\phi_{{\rm diff}}$ for each modulation. The errors in the emission line velocities were approximated as symmetric distributions, using the values summarized in Tables \ref{tab:lyas}--\ref{tab:lyafe}. In Figure \ref{fig:ccfres}, the CCF values for the best-fit velocities and distributions of $\phi_{{\rm diff}}$ are shown. In Figure \ref{fig:ccfpd}, $\phi_{{\rm diff}}$ for each element with a 90\% confidence level are shown. \par

The values of $\phi_{\rm diff}$ were $0.04 \pm 0.01$, $0.09 \pm 0.02$, $0.11 \pm 0.02$, and $0.17 \pm 0.03$ orbital period for H-like Fe, Ca, Ar, and S, respectively. All errors represent the 90\% confidence interval. 
The $\phi_{\rm diff}$ of the H-like Fe was the smallest suggesting the strongest synchronization with the motion of the compact object. As the atomic number decreases, $\phi_{\rm diff}$ increases, with H-like S being significantly larger than the other elements. \par

\section{Discussion}
\label{sec:disc}
We discuss the stellar wind structure based on the results obtained from the $XRISM$ observations. In Section \ref{subsec:pwork}, we compare our results with those obtained from the previous $Chandra$ observations. In Section \ref{subsec:cak}, we calculate the distribution of H-like ions in the Cyg X-3 system based on a spherically symmetric stellar wind model. Then, we discuss the spatial structure of the stellar wind within the Cyg X-3 system by comparing these calculations with our measurements. \par

\subsection{Comparison of the $XRISM$ results with previous studies}
\label{subsec:pwork}


We compared the modulations of emission line velocities of Cyg X-3 with the $Chandra$ observations reported by Vilhu \& Koljonen (2025), Vilhu et al. (2009), and Stark \& Saia (2003). The modulations of Fe Ly$\alpha$ lines in those studies showed little difference from our results. In contrast, the Ca, Ar, and S Ly$\alpha$ lines analyzed by Vilhu \& Koljonen (2025) and Stark \& Saia (2003) showed redshift of the baselines of the Doppler modulations, while the $XRISM$ results did not show such significant shifts of the baselines. 
Although those differences may be partly due to change of the wind conditions, we need to stress that the absorption components are newly introduced into the fitting model to explain the P Cygni profiles detected with $XRISM$. 
Vilhu et al. (2009) used a model including absorption lines for the S Ly$\alpha$ lines, but their analysis still yielded a velocity amplitude smaller than that derived from $XRISM$.
This is probably because the orbital phase was divided into only four segments due to the lack of photon statistics. Additionally, our analysis using a two-line component model of Ly$\alpha_{1}$ and Ly$\alpha_{2}$ lines improves the accuracy of the velocity measurement.
These comparisons demonstrate that the high energy resolution and large effective area of $XRISM$ enable the precise modulation analysis that is difficult to perform with the $Chandra$ data. \par

We compare the line widths, fluxes, and optical depths with previous studies. 
As suggested by Koljonen \& Maccarone (2017), the line widths may include contributions from turbulence. 
As shown in Figure \ref{fig:orb_mod}, the emission and absorption line intensities reach their minimum and maximum around phases 0.0 and 0.5, respectively. 
The similar modulations are observed in the Si Ly$\alpha$ lines with $Chandra$ (Vilhu et al. 2009).  
 \par


\subsection{Spatial distribution of H-like ions}
\label{subsec:cak}
We discuss the stellar wind structure of Cyg X-3 based on the observed modulations. For this purpose, we calculated the spatial distribution of H-like ions within the system and compared the results with the observation. We estimated the density and velocity of the wind using the CAK model (Castor et al. 1975), a spherically symmetric stellar wind model that considers line-driven acceleration by the WR star. We then calculated the spatial distribution of the ions under the condition of photoionization equilibrium.\par

\subsubsection{Photoionized stellar wind model}
\label{subsubsec:calc}
\begin{figure}[!]
\includegraphics[keepaspectratio,scale=1.0,angle=0]{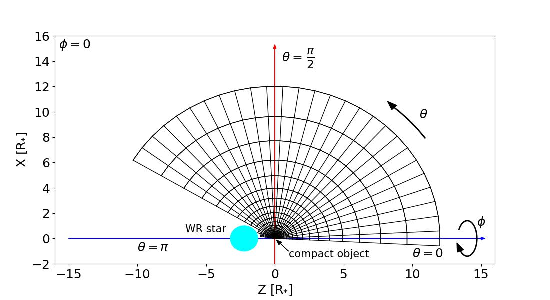}
 \caption{Grid definition for the CLOUDY calculation for $\phi = 0$. The red and blue arrows denote the X and Z axes, respectively. The compact object located at the origin.}
 \label{fig:grids}	
\end{figure}

We arrange a radial grid starting from the companion object (X, Z = 0.0) and divide it into $N$ = 32 segments as shown in Figure~\ref{fig:grids} according to the following equation, ensuring that the grid becomes smaller toward the inner regions:
\begin{equation}
    r_{{\rm grid},j} = r_{\rm min}\left( \frac{r_{\rm max}}{r_{\rm min}} \right)^{\frac{j-1}{N-1}}
\end{equation}
where $r_{{\rm grid},j}$, $r_{\rm min}$, and $r_{\rm max}$ represent the inner radius of the $j$-th grid, and the inner and outer boundaries of the radial range of interest, respectively. All radii are measured from the compact object. The outer boundary of the computational domain is set to $r_{\rm max} = 10^{12}$ cm, corresponding to the binary scale. The inner boundary is set to $r_{\rm min} = 10^{9}$ cm, since within this radius the degree of photoionization is so high for all H-like ions that their contributions to the emission and absorption lines are negligible. The angles $\theta$ and $\phi$ are defined from the Z-axis and X-axis, respectively, as typical spherical coordinates. Since the stellar wind geometry is axisymmetric about the Z-axis, the calculation was performed for $\phi = 0$ and within the range $0 \leq \theta \leq \pi$. \par

We conducted the numerical calculations of photoionization driven by X-rays from the compact object with CLOUDY 23.01 (Gunasekera et al. 2023). We arranged the grids in three-dimensional space under the definition given in Figure \ref{fig:grids}, and then conducted the one-dimensional calculation for each $\theta$. 
For the continuum spectrum from the compact object, we used the best-fit model (\texttt{constant$\times$phabs$\times$edge$_{1}\times$edge$_{2}\times$simpl$\times$(diskbb + gauss)}) obtained from the $NuSTAR$ data analysis in Appendix. 
The input source spectrum includes \texttt{diskbb} with an inner temperature of 1.5 keV and \texttt{simpl} with the photon index of 2.5 and the scattering fraction of 2.43$\times 10^{-3}$.
The WR star was placed along the Z-axis. 
No grid points were assigned in the shadowed region of the WR star, where the X-rays from the compact object are blocked by the WR star and do not reach directly, resulting in the negligibly small fractions of H-like Fe, Ca, Ar, and S ions.
\par

We modeled the plasma density in each grid based on the CAK model. 
In the CAK model, the stellar wind velocity $v_{\rm wind}$ is expressed by parameterizing the UV line-driven acceleration from the WR star with a dimensionless parameter $\beta$. 
Hereafter, we define the distance from the WR star as $r_s$.
Using the mass conservation law, the wind density $n(r_{s})$ is then calculated from $v_{\rm wind}$. The $v_{\rm wind}(r_{s})$ and $n(r_{s})$ at $r_{s}$ are expressed as:
\begin{equation}
\label{eq:vr}
    v_{\rm wind}(r_{s}) = v_{\infty}\left(1-\frac{R_{*}}{r_{s}}\right)^{\beta},
\end{equation}
\begin{equation}
\label{eq:nh}
    n(r_{s}) = \frac{\dot{M}}{4\pi m_{p}r_{s}^{2}v_{\rm wind}(r_{s})}
\end{equation}
where $v_{\infty}$, $R_{*}$, $\dot{M}$ and $m_{p}$ represent the terminal velocity of the stellar wind, the stellar radius, the mass-loss rate and the proton mass, respectively. 
We adopted $\beta$ = 1 and $v_{\infty}$ = 1600 km s$^{-1}$, which are typical values for WN-type WR stars (Lefever et al. 2023). We assumed the WR star's radius of $R_{*} = 1.5~R_{\odot}$ based on the known upper limit of $R_{*}$ < 2 $R_{\odot}$ (Koljonen \& Maccarone 2017). The uncertainty in $\dot{M}$ was large ranging between $\sim 10^{-6}$--$10^{-5}$ M$_\odot$ yr$^{-1}$ (Koljonen \& Maccarone 2017). From this range, we considered five values of mass-loss rates, 1$\times 10^{-6}$, 2$\times 10^{-6}$, 5$\times 10^{-6}$, 1$\times 10^{-5}$M$_\odot$ yr$^{-1}$, and 2$\times 10^{-5}$M$_\odot$ yr$^{-1}$, for the calculations. \par


Using the derived $n(r_{s})$, we calculated the density of each ion $n_{\rm ion}(r_{s})$ in the Cyg X-3 system using the following equation:
\begin{equation}
    n_{\rm ion}(r_{s}) = n(r_{s}) f_{\rm ion} Z_{\rm elem}
\end{equation}
where $f_{\rm ion}$, $Z_{\rm elem}$ denote the ion fraction and elemental abundance relative to hydrogen, respectively. 
We assumed $Z_{\rm elem}$ to be solar abundances as in XRISM Collaboration et al. (2024). The abundances relative to hydrogen are 2.82$\times$10$^{-5}$, 2.29$\times$10$^{-6}$, 2.51$\times$10$^{-6}$, and 1.84$\times$10$^{-5}$ for Fe, Ca, Ar, and S, respectively (Grevesse \& Sauval 1998; Holweger 2001). 
Note that assuming solar abundances is not strictly correct since the WR stellar wind has a hydrogen-depleted composition. \par

\begin{figure*}[!]
\includegraphics[keepaspectratio,scale=0.917,angle=0]{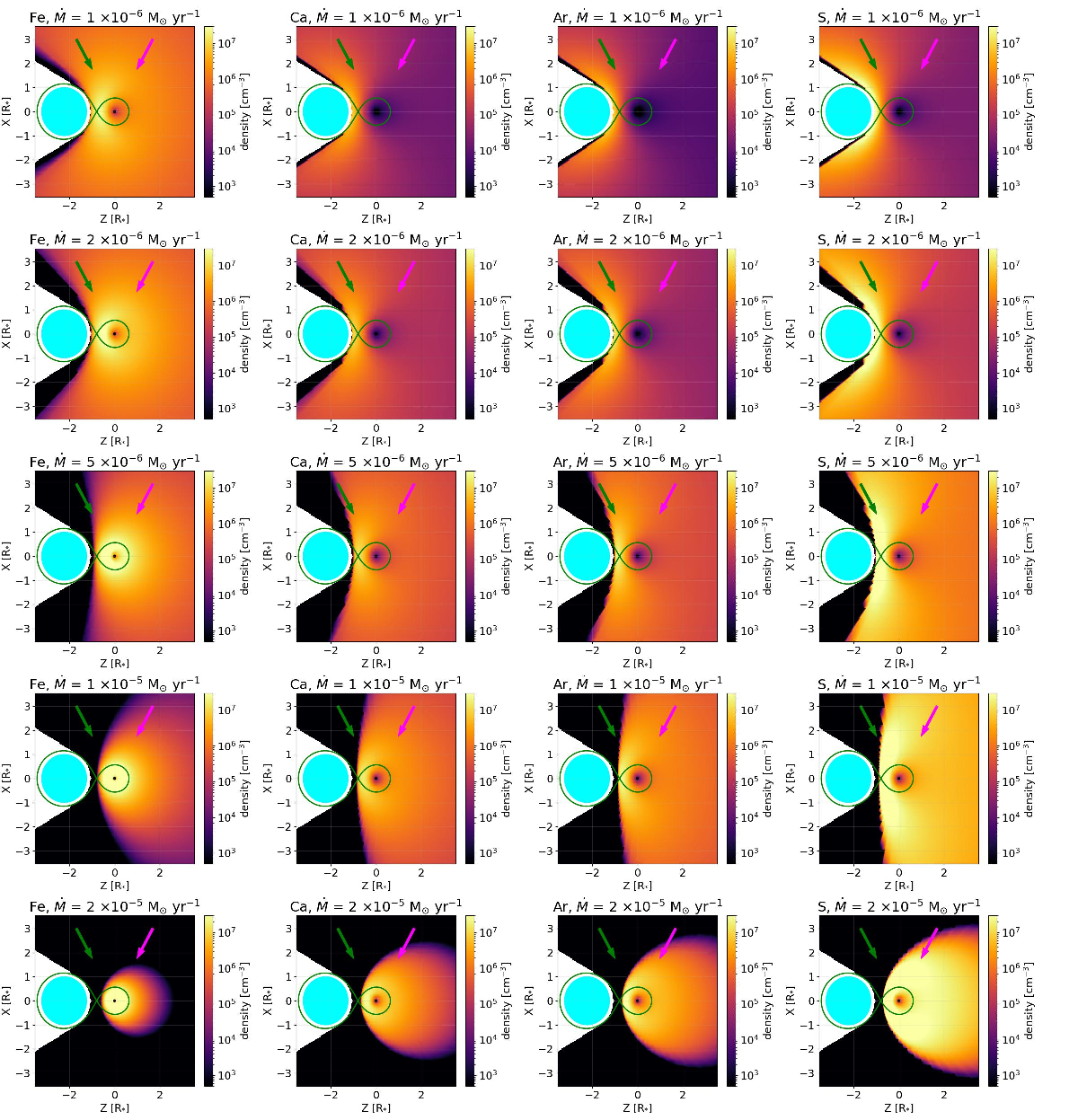}
 \caption{Calculated $n_{\rm ion}(r_{s})$ distribution on the orbital surface. From left to right, the density distributions of H-like Fe, Ca, Ar, and S are shown, and from top to bottom, the calculation results for the mass loss rates of 1, 2, 5, 10, 20$\times10^{-6}$ M$_{\odot}$ yr$^{-1}$ are presented. The green solid line represents the Roche lobe. The green and magenta arrows indicate the lines of sight at phases 0.0 and 0.5, respectively, when the system is viewed edge-on (i.e., with the X-axis interpreted as the Y-axis in this figure). Each point's value is obtained through logarithmic interpolation using the calculated values of each grid. For the purpose of logarithmic plotting, grids with a value of zero are assigned a sufficiently small value ($n_{ion}~=10^{-12}$ cm$^{-3}$) for proper visualization.}
 \label{fig:cldifrac}
\end{figure*}

Figure~\ref{fig:cldifrac} shows the calculated ion densities with each $\dot{M}$, superimposed with the Roche lobe to roughly evaluate the extent to which the wind can be captured by the gravity of the compact object.
Heavier elements tend to be more concentrated within the Roche lobe since ions with higher atomic numbers are formed in environments with higher degrees of ionization. 
For all elements, the distributions of the H-like ions tend to move from the WR star to the vicinity of the compact object as the $\dot{M}$ increases. 
This is because the higher $\dot{M}$ increases the stellar wind density, requiring proximity to the compact object to form the H-like ions. Under the same $\dot{M}$, the H-like Ca, Ar, and S show similar distributions, while the H-like Fe is more concentrated near the compact object. 
For $\dot{M} = 1 \times 10^{-6}$ and $2 \times 10^{-6} M_{\odot} \mathrm{yr}^{-1}$, H-like Fe is distributed in the region between the compact object and the WR star, rather than being concentrated within the Roche lobe.
The other ions are predominantly distributed in the region surrounding the WR star.
For $\dot{M} = 5 \times 10^{-6}$ and $1 \times 10^{-5} M_{\odot} \mathrm{yr}^{-1}$, H-like Fe is distributed within the Roche lobe. 
The other ions are distributed over a wide region of the system.
For $\dot{M} = 2 \times 10^{-5} M_{\odot} \mathrm{yr}^{-1}$, all ions are predominantly distributed within the Roche lobe.
\par


We estimate the expected line modulation based on the calculated density distributions of the H-like ions shown in Figure \ref{fig:cldifrac}.
The H-like ions are distributed on the side of the compact object as seen from the WR star, corresponding to the region with $Z > -2 R_{*}$.
It is expected that the ions responsible for emission are partly obscured by the WR star around the orbital phase of 0.0, resulting in weaker emission line flux compared to orbital phases around 0.5.
As for the absorption, Figure~\ref{fig:cldifrac} shows that, as the mass-loss rate increases, the number of ions along the line of sight at orbital phase 0.5 increases, while that at phase 0.0 decreases. 
We calculated the ratio of the column density of ions at phase 0.5 to that at phase 0.0, and summarized the results in Table \ref{tab:columndions}.  
The column densities used to derive the ratios listed in Table~\ref{tab:columndions} are obtained by integrating the ion number density from the position of the compact object, that is, (X, Z) = (0, 0).
Table~\ref{tab:columndions} shows that the column density predicted by the CAK model at phase 0.5 is smaller than or comparable to that at phase 0.0.  \par

We discuss the relationship between the stellar wind dynamics and $\phi_{\rm diff}$.
If the ions follow the motion of the compact object, then $\phi_{\rm diff} = 0.0$ (purple line in Figure~\ref{fig:orb_mod}).
As the distance from the compact object increases, the influence of the orbital motion weakens, and the velocity modulation increasingly reflects the motion of the stellar wind radially accelerated by the WR star.
Since the stellar wind is axisymmetric about the the axis connecting the compact object and the center of the WR star, we assume that the wind motion at the first Lagrange point (L$_1$ point) represents the average motion of the overall wind.
The line-of-sight velocity of the WR-star-accelerated stellar wind at the L$_1$ point without the influence of the compact object is shown with a green line in Figure~\ref{fig:orb_mod}.
If the stellar wind were unaffected by the orbital motion of the compact object and moved purely radially outward from the WR star in the extended region, the redshift would reach its maximum at phase 0.0 and the blueshift at phase 0.5, yielding $\phi_{\rm diff}$ = 0.25. \par

\begin{table}[!]
  \tbl{The ratio of the ion column densities at phase 0.5 to that at phase 0.0}{%
   \begin{tabular}{cccccc}
      \hline \hline 
      $\dot{M}$ [$10^{-6} M_{\odot}$yr$^{-1}$] & 1 & 2 & 5 & 10 & 20\\ 
      \hline 
      H-like Fe & 0.6 & 0.8 & 1.0 & 1.0 & 1.0 \\
      H-like Ca & 0.1 & 0.2 & 0.6 & 1.0 & 1.0 \\
      H-like Ar & 0.1 & 0.1 & 0.3 & 0.9 & 1.0  \\
      H-like S & 0.1 & 0.1 & 0.2 & 0.8 & 1.0 \\
      \hline \hline \\
    \end{tabular}}\label{tab:columndions}
\end{table}

\subsubsection{Interpretation of the observation results}
\label{subsubsec:cldcomp}

We interpret the velocity modulations of the emission lines observed with $XRISM$ based on our calculation results. 
As shown in Figure \ref{fig:ccfpd}, the value of $\phi_{\rm diff}$ for H-like Fe is $0.04 \pm 0.01$, which is significantly smaller than those of the other elements. 
This result indicates that H-like Fe follows the compact object.
In contrast, the value of $\phi_{\rm diff}$ for H-like S is $0.17 \pm 0.03$, which is significantly closer to 0.25 than those of the other elements. 
The value of $\phi_{\rm diff}$ for the S Ly$\alpha$ line suggests that H-like S is distributed farther from the compact object and follows the radial stellar wind motion as described in the final paragraph of Section~\ref{subsubsec:calc}.
Figure \ref{fig:cldifrac} shows that H-like S is more widely distributed farther from the compact object than H-like Fe.
The H-like ions of Ca and Ar are considered to have intermediate behavior between Fe and S since the $\phi_{\rm diff}$ of the Ca and Ar Ly$\alpha$ lines are larger than that of the Fe Ly$\alpha$ lines but smaller than that of the S Ly$\alpha$ lines.
The observed velocity amplitude of the Fe Ly$\alpha$ line is comparable to that of the orbital motion of the compact object. 
The velocity amplitudes of the Ca, Ar, and S Ly$\alpha$ lines are approximately $500~\mathrm{km~s^{-1}}$, consistent with the stellar-wind velocity in the emission region. 
A representative emission region for these lines is the vicinity of the L$_1$ point.  \par

We compare the results of our calculations with the observed modulations of the emission-line flux and the absorption, which reach their maxima and minima around the orbital phases of 0.5 and 0.0, respectively.
The modulations of the emission-line intensities are interpreted as reflecting partial occultation by the WR star.
As described in section \ref{subsubsec:calc}, the calculated results shown in Table~\ref{tab:columndions} indicate that the absorption at phase 0.5 never exceeds that at phase 0.0. 
The observational results cannot be explained by the calculated results alone.
The column densities in Table~\ref{tab:columndions} correspond to absorption of the continuum emission, whereas the observed spectra include absorption of line emission. 
The observed column density for the line emission represents a superposition of column densities from different locations since the line-emitting region is spatially extended rather than point-like.
The column density at phase 0.5 can exceed that at phase 0.0 depending on the choice of the starting point of the column.
This effect may contribute to the discrepancy between the assumed models and the observations.
Furthermore, the fact that the absorption observed at phase 0.5 is stronger than that at phase 0.0 may suggest the presence of additional absorption at phase 0.5. \par

As shown in Figure \ref{fig:orb_mod}, the peak of the absorption does not exactly coincide with phase 0.5.
Such an asymmetric modulation cannot be explained by the CAK model.
Similar asymmetric variations of the absorption are observed in several HMXBs, and are interpreted as inhomogeneous structures crossing the line of sight, such as accretion wake (Abalo et al. 2024; Odaka et al. 2013; Blondin et al. 1990).
The accretion wake is thought to have a spatial scale comparable to the stellar radius (Malacaria et al. 2016). 
Assuming the stellar-radius-scale wake is present in Cyg X-3, it may contribute not only to the Fe Ly$\alpha$ absorption lines but also to the modulation of the S Ly$\alpha$ absorption lines, which originate from a relatively extended region.
An additional absorption component at orbital phase 0.5 that is not accounted for in the CAK model can potentially be interpreted as the accretion wake, which is denser than the surrounding stellar wind.
In addition to the accretion wake, dense clumps compressed by a bow shock have also been theoretically proposed as a possible cause of asymmetric absorption variability (Vilhu \& Hannikainen 2013). \par


We discuss the observed line widths using the calculation results. 
The electron temperature within the system obtained by the calculation is below 1 keV, which corresponds to the thermal velocity width of less than 100 km s$^{-1}$ assuming a Maxwellian distribution.
The observed line widths shown in Figure \ref{fig:orb_mod} are significantly larger than the calculated thermal velocity. 
The velocity widths contain a contribution from the stellar wind accelerated along the radial streamlines from the WR star.
In addition, the velocity dispersion of wind falling onto the compact object from various directions may also contribute. 
This contribution is expected to be particularly significant for ions near the compact object, such as H-like Fe. \par

We estimate the mass-loss rate based on the comparison between the calculations and the observations.
As already suggested by the observed Doppler shift modulations, the H-like Fe ions are concentrated in the vicinity of the compact object, while H-like S ions are distributed across the wider region in the system, extending to the compact object side as seen from the WR star.
These observationally  inferred distributions are qualitatively well reproduced by the density distribution calculations for mass-loss rates of $5\times10^{-6}$ and $1\times10^{-5}\,M_\odot\,\mathrm{yr}^{-1}$  (see in Figure~\ref{fig:cldifrac}).
\par

For more quantification of the stellar wind structure, it will be necessary to consider not only line-driven acceleration by the WR star but also various physical effects such as the gravity and radiative pressure of the compact object, and suppression of UV acceleration due to the X-ray-driven photoionization. Since it is difficult to model these effects analytically, hydrodynamical and radiative transfer simulations will be required for future studies. Our results demonstrate that high-resolution spectroscopy with $XRISM$ has great capability of probing the stellar wind structure, representing the milestone toward this goal. \par

\section{Conclusion}
\label{sec:conc}
We measured the orbital modulations of the Fe, Ca, Ar, and S Ly${\alpha}$ lines. We extracted spectra for eight orbital-phase bins. The extracted spectra showed the detailed line structures, such as the P Cygni profiles. We performed spectral fitting for Fe, Ca, Ar, and S Ly${\alpha}$ lines using the models including both the emission and absorption lines, which was made possible by the high spectral resolution and large effective area of $XRISM$. The phase differences between the Doppler shifts of the Fe, Ca, Ar, and S Ly${\alpha}$ emission lines and the orbital motion of the compact object were 0.04, 0.09, 0.11, and 0.17 in units of  the orbital period, respectively. The amplitudes of the Doppler shifts were approximately 500 km s$^{-1}$ for all lines. The widths of those lines showed the orbital modulations, and ranged from 400 to 1000 km s$^{-1}$. The intensities of both the emission and absorption lines reached their minima and maxima near the phase 0.0 and 0.5, respectively. \par

We interpreted the observed line modulations based on the calculation of the ion distributions in the WR stellar wind photoionized by the X-rays from the compact object. We find that the modulation in the Doppler shift of the Fe Ly${\alpha}$ lines reflects the motion following the compact object most strongly among all elements, whereas the modulation in the S Ly${\alpha}$ lines reflects the radially accelerated stellar wind from the WR star more strongly than  the Fe Ly${\alpha}$ lines. 
This result indicates that H-like Fe is distributed in the vicinity of the compact object whereas H-like S is distributed across the wider region in the system.
The Ca and Ar Ly$\alpha$ lines have intermediate behavior between  the Fe and S Ly$\alpha$ lines.
From these distribution characteristics, we estimate that the mass loss rate of the stellar wind is approximately $5 \times 10^{-6}$--$1 \times 10^{-5}\ M_{\odot}\ {\rm yr}^{-1}$.
The widths of lines are reasonably understood by the velocity dispersion of the stellar wind, which is radially accelerated by the WR star, in addition to thermal broadening.
The width of the Fe Ly$\alpha$ lines may also be influenced by the motion of the stellar wind falling onto the compact object.
The modulations of the emission line intensities indicate that the WR star partially obscures the line emission region of the stellar wind. 
The asymmetric modulation of the absorption around phase 0.5 suggests the presence of inhomogeneous structure possibly associated with an accretion wake.
This structure may contribute to the enhanced absorption around orbital phase 0.5. \par

\begin{ack}
T.H. is financially supported by the JST SPRING, grant number JPMJSP2138. This work is supported by JSPS KAKENHI grant numbers JP22H00128, JP22K18277, and JP22H00158, and JST SPRING, grant number JPMJSP2108. We gratefully acknowledge all those who contributed to the $XRISM$ mission.
\end{ack}

\appendix
\section*{$NuSTAR$ data}
\label{app:nustar}
We analyzed the spectra observed with $NuSTAR$ to construct the input radiation spectrum table from the compact object for the photoionization equilibrium calculations in Section \ref{subsec:cak}. The purpose of this analysis is not to carry out a detailed physical discussion of the $NuSTAR$ spectra, but rather to reproduce the overall spectral shape. We did not perform rigorous statistical tests, and our evaluation was limited to assessing the level of residuals compared to the data. \par

\subsection{Observation and data reduction}

For the photoionization equilibrium calculations (see Section \ref{subsec:cak} for details), an input radiation spectrum from the compact object which serves as the ionizing source is required. The X-rays from the compact object are considered hard because they can ionize iron. Data obtained with an instrument that has a large effective area in the hard X-ray band above 7 keV are most suitable to determine the radiation spectrum from the compact object. For this purpose, observational data from $NuSTAR$ are useful. $NuSTAR$ is one of the X-ray observatories that have large effective areas in the hard X-ray band. \par 

$NuSTAR$ conducted coordinated observation of Cyg X-3 with $XRISM$. The observation was conducted from 2024 March 24 00:16:09 TT to 2024 March 24 22:16:09 TT, resulting in the net exposure time of 25895 seconds. In Figure \ref{fig:xrism_lc}, the period during which the NuSTAR observations were performed is plotted on the XRISM/Resolve light curve. We extracted and analyzed the time averaged spectra of $NuSTAR$/FPMA and FPMB. We used NuSTARDAS version 2.1.2 and XSPEC 12.4.0b in Heasoft 6.33.1 to extract and analyze spectra, respectively. All options were set to their default values (refer to the NuSTAR Data Analysis Software Guide) during data extraction with NuSTARDAS.\par

\begin{figure}[!]
 \begin{minipage}{0.1\hsize}
\includegraphics[keepaspectratio,scale=0.959,angle=0]{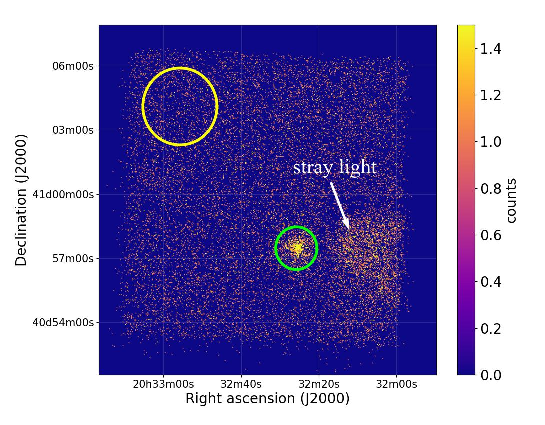}
 \end{minipage}
 \caption{Observed image with $NuSTAR$/FPMA in the energy band of 20--79 keV. The extended structure observed immediately to the right of Cyg X-3 is the stray light. The green and yellow circles represent the region from which we extracted spectra of Cyg X-3 and background, respectively.}\label{fig:nusao}
\end{figure}

 We extracted the spectra of Cyg X-3 and the background from circular regions with radii of 1 arcmin and 1.8 arcmin, centered at (RA, Dec) = (\timeform{20h32m25.78s}, \timeform{+40D57'27.90''}) and (\timeform{20h32m55.75s}, \timeform{+41D04'06.09''}), respectively. The stray light was present near Cyg X-3, and thus the radius of the extraction region was adjusted to avoid including it. The stray light and the regions used for the analysis are shown in Figure \ref{fig:nusao}. We also used NuSTARDAS version 2.1.2 with the CALDB version 20240715 to make the response files. \par

\subsection{Analysis and results}
\label{sec:nuana}
We fitted the spectra observed with $NuSTAR$. We used the spectra in the 3-20 keV range, as the count rates above 20 keV were comparable to the background. All model parameters were linked between FPMA and FPMB, with a scaling factor applied to FPMB (\texttt{constant}) to account for a 5\% systematic calibration difference in normalization between the two modules (Harrison et al. 2013). The spectral fitting was carried out using the C-statistic. We initially fitted the spectra using an absorbed multicolor disk blackbody model (\texttt{phabs$\times$diskbb}) alone. \texttt{phabs} and \texttt{diskbb} represent the photoelectric absorption within the Cyg X-3 system and interstellar medium, and the radiation from the accretion disk, respectively. This analysis resulted in residuals: edge-like features in the 7-8 keV and 8-9 keV ranges, an emission-line-like excess around 7 keV, and an overall excess above 10 keV. To address these features, we added two absorption edges (edge${1}$ and edge${2}$) and a Gaussian emission component. We attributed the excess above 10 keV to inverse Compton scattering by electrons, and convolved \texttt{diskbb} with the Comptonization model (\texttt{simpl}) to phenomenologically account for it. The photon index of the \texttt{simpl} component, $\Gamma$, was fixed at 2.5 (Koljonen et al. 2018), as it could not be constrained using only the narrow high-energy band above 10 keV. The Gaussian line width was fixed to zero because $NuSTAR$ lacks the energy resolution to resolve stellar wind velocity broadening. The final model used was:
\texttt{constant$\times$phabs$\times$edge$_{1}\times$edge$_{2}\times$simpl$\times$(diskbb + gauss)}. \par

\begin{figure}[!]
\includegraphics[keepaspectratio,scale=1.042,angle=0]{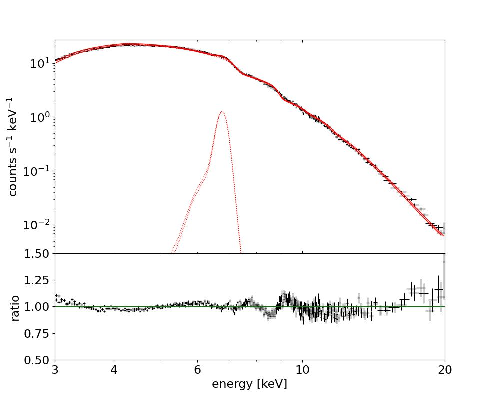}
 \caption{Observed spectra by $NuSTAR$. The upper and lower parts show the observed spectrum and the residual, respectively. The black crosses represent the observed spectra by $NuSTAR$/FPMA and FPMB. The red solid lines represent the best-fit model. The red dotted lines represent each component.}
 \label{fig:nuana}
\end{figure}

Figure \ref{fig:nuana} and Table \ref{tab:nuspec} show the results of the spectral fitting. We found edge${1}$ and edge${2}$ to be at energies of 7.3 keV and 8.9 keV, respectively, which we interpret as the K-edges of neutral Fe and He-like Fe. 
The emission line represented by the Gaussian function is the merged feature of the He-like and H-like Fe K$\alpha$ lines. As a result of the fitting, although the ratio of residuals to the data exceeded several tens of percent at maximum, the ratios were less than 10\% in 70\% of the bins. Therefore, we adopted the best-fit model as the source spectrum for the calculations in Section \ref{subsubsec:calc}. For the calculation, we used the best-fit model shown in Table \ref{tab:nuspec} as the ionized source spectrum. In this model, the portion corresponding to the intrinsic radiation of the compact object is \texttt{diskbb} convolved with \texttt{simpl}. In the calculation, we used only the part consisting of \texttt{diskbb} and \texttt{simpl} as the input spectrum. \par

\begin{table}[!]
  \tbl{The spectral fitting results observed with $NuSTAR$}{%
   \begin{tabular}{cccc}
      \hline
      \hline
      Model & Parameter & value \\ 
      \hline
      phabs & $N_{H}$ [10$^{22}$ cm$^{-2}$] & 6.40$^{+0.07}_{-0.08}$ \\
      \hline
      edge$_{1}$ & energy [keV] & 7.28$\pm0.01$ \\
         & depth & 0.322$\pm0.005$ \\
      \hline
      edge$_{2}$ & energy [keV] & 8.89$\pm0.01$ \\
         & depth & 0.488$\pm0.008$ \\
      \hline
      simpl & scattering fraction [10$^{-3}$] & 2.43$\pm0.29$ \\
      \hline
      diskbb & $T_{in}$ [keV] & 1.539$^{+0.006}_{-0.005}$ \\
            & Norm & 116$\pm2$ \\
      \hline
      gaussian & energy [keV] & 6.760$^{+0.007}_{-0.002}$ \\
            & flux [10$^{-3}$ photons cm$^{-2}$ s$^{-1}$] & 2.62$\pm0.15$ \\
      \hline
    constant & factor & 1.000$\pm0.002$ \\
      \hline
     & flux\footnotemark[$*$]  [10$^{-9}$ erg cm$^{-2}$ s$^{-1}$] & 5.09$\pm0.01$ \\
      \hline
      \hline
    \end{tabular}}\label{tab:nuspec}
\begin{tabnote}
\footnotemark[$*$] The absorbed X-ray flux in 2-100 keV band. \\
Note: The errors are 90 \% confidence interval. $C$ and degree of freedom are 2493.78 and 837, respectively.\\
\end{tabnote}
\end{table}


\begin{thebibliography}{}

\bibitem[bib0]{}
Abalo, L., et al. 2024, \aap,  692, A188

\bibitem[bib1]{}
Antokhin, I. I., Cherepashchuk, A. M., Antokhina, E. A., \& Tatarnikov, A. M. 2022, \apj, 926, 123

\bibitem[bib1p1]{}
Antokhin, I. I., \& Cherepashchuk, A. M., Antokhina, E. A. 2019, \apj, 871, 244

\bibitem[bib1p1p5]{}
Blondin, J. M., Kallman, T. R., Fryxell, B. A., \& Taam, R. E. 1990, \apj, 356, 591

\bibitem[bib3]{}
Cash, W. 1979, \apj, 228, 939

\bibitem[bib4]{}
Castor, J. I., Abbott, D. C., \& Klein, R. I. 1975, \apj, 195, 157

\bibitem[bib5]{}
Grevesse, N., Sauval, A. J. 1998, \ssr, 85, 161

\bibitem[bib5p5]{}
Gu, M. F. 2008, Canadian Journal of Physics, 86(5), 675

\bibitem[bib6]{}
Gunasekera, C. M., van Hoof, P. A. M., Chatzikos, M., \& Ferland, G. J. 2023, RNAAS, 7, 246

\bibitem[bib6p25]{}
Harrison, F. A., et al. 2013, \apj, 770, 103

\bibitem[bib7]{}
Holweger, H. 2001, \aip 598, 23

\bibitem[bib8]{}
Kallman, T., et al. 2019, \apj, 874, 51

\bibitem[bib10]{}
Koljonen, K. I. I., \& Maccarone, T. J. 2017, \mnras, 472, 2181

\bibitem[bib10p5]{}
Koljonen, K. I. I., Maccarone, T., McCollough, M. L., Gurwell, M., Trushkin, S. A., Pooley, G. G., Piano, G., \& Tavani, M. 2018, \aap, 612, A27

\bibitem[bib11]{}
Lefever, R. R., Sander, A. A. C., Shenar, T., Poniatowski, L. G., Dsilva, K., \& Todt, H. 2023, \mnras, 521, 1374

\bibitem[bib11p5]{}
Malacaria, C., Mihara, T., Santangelo, A., Makishima, K., Matsuoka, M., Morii, M., \& Sugizaki, M. 2016, A\&A, 588, A100

\bibitem[bib12p5]{}
Miura, D., et al. 2025, \pasj, 77, S86

\bibitem[bib13]{}
Odaka, H., Aharonian, F., Watanabe, S., Tanaka, Y., Khangulyan, D., \& Takahashi, T. 2011, \apj, 740, 103

\bibitem[bib13p50]{}
Odaka, H., Khangulyan, D.,  Tanaka, Y. T., Watanabe, S., Takahashi, T., \& Makishima, K. 2013, \apj, 767, 70

\bibitem[bib13p51]{}
Perri, M., et al. 2021, "The NuSTAR Data Analysis Software Guide Version 1.9.7"

\bibitem[bib13p52]{}
Reid, M. J., Miller-Jones, J. C. A. 2023, \apj, 959, 85

\bibitem[bib15]{}
Stark, M. J., \& Saia, M. 2003, \apj, 587, L101

\bibitem[bib16]{}
Tashiro, M., et al. 2025, \pasj, 77, S1

\bibitem[bib16p5]{}
Vilhu, O., \& Koljonen, K. I. I. 2025, \aap, 699, A270

\bibitem[bib17]{}
Vilhu, O., \& Hannikainen, D. C. 2013, \aap, 550, A48

\bibitem[bib18]{}
Vilhu, O., Hakala, P., Hannikainen, D. C., McCollough, M., \& Koljonen, K. 2009, \aap, 501, 679

\bibitem[bib18p5]{}
Watanabe, S., et al. 2006, \apj, 651, 421

\bibitem[bib2]{}
XRISM Collaboration, et al. 2024, \apjl, 977, L34

\bibitem[bib19]{}
Zdziarski, A. A., Miko\l{}ajewska, J., \& Belczy$\acute{\rm n}$ski, K. 2013, \mnras L, 429, L104


\end{thebibliography}
\end{document}